
\input harvmac

\font\names=cmbx10 scaled\magstep1
\baselineskip=20pt
\Title{PUP-TH-1374 (1993)}
{\vbox{\centerline
{Family Symmetry, Fermion Mass Matrices}
 \centerline{ and Cosmic  Texture}}}

\font\names=cmbx10 scaled\magstep1
\centerline{\bf\names Michael Joyce}
\centerline{ and }
\centerline{\bf\names Neil Turok}
\centerline{Joseph Henry Laboratories}
\centerline{Princeton University, Princeton NJ08544.}
\baselineskip=12pt
\centerline{\bf Abstract}
\noindent
The observed replication of fermions in three families
is undoubtedly a reflection of a deeper symmetry underlying the
standard model. In this paper we investigate one very
elementary possibility,  that physics above the grand unification
scale is
described by the symmetry group
 $G\times SU(3)_{\rm fam}$ with $G$ a gauged grand unified
 group, and $SU(3)_{\rm fam}$ a {\it global} family symmetry.
The breaking of this symmetry at the GUT scale produces
global texture, providing a
mechanism for structure formation  in the universe, and sets
strong constraints on the
low energy fermion mass matrix.
With the addition of a ${\bf 45}$ Higgs and certain assumptions
about the relative strength of Higgs couplings, the simplest
$SU(5)$ theory yields an eight
parameter
 form for the fermion mass matrices, which we show is consistent
with the thirteen observable masses and mixing angles.
 We discuss the natural suppression of
 flavour changing neutral currents
 (FCNCs) and emphasise the rich low energy Higgs sector.
With minor assumptions,
the theory  unifies consistently with the recent LEP data.
We consider the extension to $G= SO(10)$, where some simplification
and further predictiveness emerges.
Finally we discuss the dynamics of $SU(3)_{\rm fam}$ texture and show
that the known constraints on unification `predict' a GUT
symmetry breaking scale of the order
required for cosmic structure formation, and
by the recent detection  of cosmic microwave anisotropy by COBE.

\Date{1/93} 

\baselineskip=24pt
\centerline{\bf 1. Introduction}

    The idea that the inhomogeneities required for structure formation
in the early universe were generated when a
symmetry of nature was broken is one of the most attractive in cosmology.
The predictions of  such theories depend on the pattern of
broken symmetry. One generic possibility involves
 an unstable defect known as global
texture \ref\turok{N. Turok, Phys. Rev. Lett., {\bf 63} (1989) 2625.}, formed
whenever a continuous nonabelian global symmetry is completely broken,
     so
that the vacuum manifold has a non-trivial
third homotopy group $\Pi_{3}(M_0)$. The breaking scale, the single
tunable parameter in the simplest texture model, is required
to be of order $ 10^{16}$ GeV
in order to fit the requirements
of structure formation, and to fit the level of microwave background
anisotropy recently detected by COBE \ref\Smoot{G. Smoot {\it et. al.},
Ap. J. {\bf 396} (1992) L1.}, \ref\BR{D.P. Bennett
and S.H. Rhie, Livermore preprint UCRL-JC-111244 (1992).},
\ref\PST{U. Pen, D. Spergel and N. Turok,
Princeton  preprint PUP-TH-1375 (1993).}.
 This coincidence with
the GUT scale is striking. This paper is devoted to
an investigation of
some of the phenomenological implications of imposing a global family
(or ``horizontal'') symmetry on a typical grand unified  theory of particle
interactions. In particular we shall show that these theories can fit the
observed low energy phenomenology and do indeed naturally
produce the  symmetry breaking scale required for cosmic structure
formation and to fit  the recent COBE results.

    The effects of global family
    symmetries have been discussed at considerable length
in the literature
\ref\reiss{D. Reiss, Phys. Lett. {\bf 115B}
(1982) 217; G. B. Gelmini, S. Nussinov and
T. Yanagida, Nuc. Phys. {\bf B219} (1983) 31; F. Wilczek, Phys. Rev. Lett
{\bf 49} (1982) 1549;
 D. Chang, P. Pal and G. Senjanovic, Phys. Lett. {\bf 153B} (1985) 407. }.
 The Goldstone bosons produced when a global symmetry is broken
(known as ``familons'' in the case of a family symmetry) have couplings to the
fermions which are inversely proportional to the breaking scale $\phi_o$.
The tightest constraints on this scale for a family symmetry come from the
FCNCs  mediated by the familons ($\phi_0>10^{10}$GeV). The constraints from
the long range forces mediated by such massless particles are in fact very
weak ($\phi_0>10^{2}$GeV) because the couplings to the fermions are  derivative
in the familon fields.

	Consider the Yukawa couplings in minimal $SU(5)$
\eqn\ea{\eqalign{ Y_{ij} \psi_i^{\alpha T}C\chi_{j \alpha \beta } H_{\beta}^*
              +  {Y'} _{ij} \varepsilon^{\alpha \beta  \gamma \delta
\epsilon } \chi_{i \alpha \beta}^T C\chi_{j \gamma \delta} H_{\epsilon}  + h.c.
     \cr}}
where $\psi$ and $\chi$ are the fermion $\overline{{\bf 5}}$ and {\bf 10}, $H$
is the Higgs
{\bf 5}, $i,j$ are family labels and C is the charge conjugation matrix. With
$<H_\alpha> = v \delta _\alpha ^5$ the first term
produces the mass matrices for the $(d,s,b)$ quarks and the charged
leptons $(e,\mu ,\tau)$ and the second the $(u,c,t)$ mass matrix. If we now
postulate a family symmetry, e.g. with the quarks in {\bf 3}s of $SU(3)$, the
nine Yukawa couplings in each term are reduced to a single one.
The question of mass
relations and mixing angles is therefore a question about what vevs
the Higgs potential can produce. In this paper we shall investigate this
in some detail.

	Such a potential is a function of many variables constrained by
the symmetries to just a few terms. We expect this to lead to constraints
on the form of the minimum of the potential. Michel's conjecture
\ref\li{L. Li, Phys. Rev. {\bf D9} (1974)
1723; J .Kim, Nuc. Phys. {\bf B196} (1982) 285.} states
that a potential involving a single irreducible representation of a
group has a minimum which leaves unbroken a maximal subgroup. A
well known example is the breaking of $SU(5)$ using the hermitian traceless
{\bf 24}. The minimum must be at one of four definite values (up to
 $SU(5)$ rotations). We are only free to adjust the overall magnitude which
is a function of the parameters in the potential. For several representations
 however there is no general result:
one must examine the particular case to see what form the vev takes.

     Below we will consider models in which we use a set of GUT scale Higgs
fields to break $SU(5)\times G_{fam}$ down to
$SU(3)_C \times SU(2)_W \times U(1)_Y$. These heavy Higgs fields couple to
the electroweak Higgs fields which give mass to the fermions.
We thus consider an
effective low energy Higgs potential which we get by substitution of
the GUT scale vevs in the total potential. This effective low energy
potential is not
completely general because of the constraint of renormalisability imposed
on the underlying theory. As a simple illustration consider
a theory with two complex fields $\phi$ and $\eta$
with the same U(1)
charge. A renormalisable potential contains no terms in $\phi^3$: thus
if $\eta$ aquires a vev, and breaks the symmetry,
the tree level effective potential for $\phi$ is not completely general.
Quantum corrections do generate the `missing'
terms, but they
 are always proportional to powers of the non-trivial couplings between the
electroweak and GUT scale fields. This follows to all orders in
perturbation theory because in the absence of these couplings
the theory is invariant under (global) transformations of the $\phi$ and
$\eta$ fields separately. It is technically `natural' to fine tune
the $\phi$-$\eta$ couplings to be small.

    This means that the quantum corrections do not spoil the
hierarchy imposed at tree level. The usual
hierarchy problem of GUTs is of course still present, and
shall not be addressed in this paper. This problem arises because
representations must be split between the
two mass scales (e.g. the Higgs ${\bf 5}$ must have an electoweak
doublet component and a superheavy proton decay mediating color triplet
component). This is
achieved by fine tuning linear combinations of cross couplings which
are not individually fine-tuned. This splitting of the gauge
representations can be achieved however with couplings which leave the
larger family symmetry on the two mass sectors intact. Provided we
not wish to split the family replicas of given gauge components we
do not escalate the hierarchy problem of the minimal GUT by adding a
family symmetry as we have described.
We shall show that with the tree
level potential alone, an acceptable form for the fermion mass
matrices is obtainable in the context of a simple
perturbative scheme. It would certainly be interesting to extend this work to
supersymmetric theories where there is some hope of explaining
the electroweak -GUT heirarchy.

     One way of phrasing the question we address is this: Does a family
symmetry broken at GUT scale leave any mark on the low energy physics? The
answer is twofold. The low energy physics reflects the family symmetry
in that while the effective Higgs potential is not invariant under
family symmetry it is not completely general with respect to
 the family indices.
The low energy fermion masses and mixing angles may or may not reflect the
family symmetry. If the Higgs potential is such
that it can produce any mass matrices by suitable choice of couplings
then we cannot exclude the possibility of such a symmetry existing
from  low energy phenomenology alone.
If on the other hand it produces mass matrices
which are predictive and fit
the data we may indeed be able to see a `shadow' of this GUT scale
symmetry without knowing more of the Higgs sector. With the increasingly
accurate measurement of the KM matrix, it seems an opportune time
to make a detailed investigation of these issues.

The outline of the paper is as follows. In Section 2 we
argue that the simplest choice for a family symmetry is $SU(3)$,
and further that this {\it must} be a {\it global} rather than
{\it gauged} symmetry for quantum consistency. In Section 3
we perform a detailed minimisation of the potential for the
$SU(5)\times SU(3)_{\rm fam}$ symmetry with a simple choice of Higgs
representations and show how mass matrices of the sort required by
phenomenology are generated in such a
potential. While this is hardly a stunning success, it
indicates that the idea of a family symmetry is viable, and gives
us some idea as to how the imposition of further symmetries might
make the theory predictive. In Section 4 we discuss flavour changing
neutral currents (FCNC). In Section 5 we discuss the running of the
tree level mass matrices, and in Section 6 a simple extension of the
$SU(5)$ theory to include a {\bf 45}, which is needed to obtain the
correct fermion masses and makes the theory consistent with the unification
of couplings. It also cures the FCNC problem of the minimal theory.
In section 7 we fit a particular ansatz to the data to demonstrate the
detailed viability of this extended model.
In Section 8 we discuss embedding the theory in $SO(10)$, and point out
how the observed mass matrices might naturally emerge in the resultant theory.
Finally, in Section 9 we discuss the cosmological implications
of $SU(3)_{\rm fam}$ texture, and relate the GUT scale to the
level of cosmological fluctuations recently detected
by COBE. In Section 10 we point out directions for  future work.

\centerline{\bf 2. Choice of Group and Fermion Representations}

    We choose to take $G_{fam}$ as a simple Lie group. The breaking
  of a
 discrete family symmetry group would produce the familiar cosmological
 domain wall problem, unless the symmetry were anomalous
 \ref\Pre{J. Preskill, S. Trivedi, F. Wilczek and M. Wise,
Nuc. Phys. {\bf B363} (1991) 207.},
in which case one might
question the logic of
imposing  a symmetry at the classical
 level when it is broken anyway by quantum effects.
 We choose further to
constrain the possible $SU(5)$ fermion representations
by insisting that we can embed  the
model in $SO(10)$ with its attractive feature that
all the fermions fit in a single ${\bf 16}$. Since
the $SU(5)\times G_{\rm fam}$ decomposition is
\eqn\edd{\eqalign{({\bf 16},{\bf 3})=({\bf 1},{\bf 3})
+({\bf 10},{\bf 3})+(\overline{\bf 5},{\bf 3})}}
the representations of the fermions are then
specified.

The only simple groups with three dimensional representations
are $SU(2)$ and $SU(3)$. $SU(3)$ has several features which make us favour it:

(i) Consider again the Yukawa couplings in \ea. Taking the assignments of
$\edd$ for the {\bf 10} and $\overline{\bf 5}$ we see that if the Higgs {\bf 5}
 is to give the fermions a mass, it must be in the following representations
\eqn\ed{\eqalign{
                 SU(2) :  {\bf 1},{\bf 3}, {\bf 5}
                 &\quad({\rm first}
                 \quad{\rm term}) \cr
                 {\bf 1}, {\bf 5}
          &\quad({\rm second}\quad{\rm term}) \cr
                 SU(3) :   \quad\overline{\bf 3},{\bf 6}&\quad({\rm first}
                 \quad{\rm term}) \cr
          \overline{\bf 6}&\quad({\rm second}\quad{\rm term}) \cr}}
 Note that the $({\bf 5},{\bf 1})$ and
$({\bf 5},{\bf 5})$ under $SU(5) \times SU(2)$ couple
to {\it both} terms whereas this does not happen for the $SU(5)\times SU(3)$
case, because the relevant representations of $SU(3)$ are complex.
This has the beneficial effect of suppressing
 FCNCs which we will argue happens quite
naturally in the simplest $SU(3)$ model partly because of this feature.

(ii) We will see below that the zero-order form of the $SU(3)$ ${\bf 6}$ and
$\overline{{\bf 6}}$ vevs (by ``zero-order'' we mean their values when
different
representations are not coupled) reproduces that of all three
fermion mass matrices
\eqn\ee{\eqalign{ \pmatrix{0  &0  &0 \cr
	                   0 &0  &0 \cr
	                   0  &0  &M \cr }}}
i.e. there is one fermion in each triplet
much heavier than the others. To produce this approximate
form in $SU(2)$ requires a fine tuning between the singlet and the {\bf 5} as
the latter is traceless
\ref\wilzchek{A gauged $SU(2)$ family symmetry
is considered in F. Wilczek and A. Zee, Phys.
 Rev. Lett. {\bf 42} (1979) 421.}.
It is not difficult to see that this also means that
the minimal number of Higgs we need in $SU(2)$ is three ({\bf 1},{\bf 3},{\bf
5}) to have any
chance of fitting the mass matrices. Even then we have 5 independent Yukawa
couplings as opposed to only 2 in the minimal $SU(3)$ model which we will
consider.

(iii) The $SU(3)$ imposed on these GUTs is anomalous because of its simple
chirality. If all fifteen left handed  fermions are placed in {\bf 3}'s
of $SU(3)$, it  is easy to check that Tr($T^a \{T^b, T^c\}$) is nonzero
so the $SU(3)$ symmetry, if gauged,
would be anomalous.
In contrast,
the {\bf 3} of $SU(2)$ is a real representation so the generators are
antisymmetric, and Tr($T^a \{T^b, T^c\}$) is zero
\ref\zoupanos{A gauged non-chiral $SU(3)$ extension of the standard model is
discussed in Zoupanos, Phys. Lett. {\bf B115} (1982) 221.}.
 Thus $SU(3)$ in contrast to $SU(2)$ cannot be gauged. This might traditionally
 be regarded as a flaw, but with the production of cosmic texture in
 mind, we regard it as a virtue - the $SU(3)$ family symmetry must be global.
Note also that there is no anomalous coupling between the Goldstone boson
of $SU(3)$ and the QCD gauge fields, because the generators of $SU(3)$
are traceless. So the $SU(3)$ Goldstone bosons remain exactly massless.

    So consider $SU(5)\times SU(3)_{fam}$. We take these representations:
\eqn\eh{\eqalign{ \Phi_{1 \alpha a b}\quad ({\bf 5},{\bf 6})
\qquad {\Phi_{2\alpha}}^{a b}    \quad ({\bf 5},\overline{\bf
6})\qquad {{\Sigma_\alpha^1}^\beta}_a\quad ({\bf 24},{\bf 3})\qquad
   {{\Sigma_\alpha^2}^\beta}_a\quad ({\bf 24},{\bf 3}) }}
$\Phi_1$ and $\Phi_2$ are the electroweak Higgs fields which produce
the mass matrices. We take two ({\bf 24},{\bf 3})s as we must completely break
$SU(3)_{fam}$ at the GUT scale ( it is easy to see that a single
 ({\bf 24},{\bf 3})
cannot achieve this). We will make the two line up to be
perpendicular in family space.

    Our mass matrices will be symmetric. This in itself is not a restriction
on the resultant fermion mass phenomenology as any mass matrices related by the
transformation
\eqn\ei{\eqalign{ M_d \rightarrow TM_dS_1^\dagger \qquad
                  M_u \rightarrow TM_uS_2^\dagger }}
where $T,S_1$ and $S_2$ are unitary matrices produce the same masses
and KM parameters. In particular the matrices can always be
made symmetric by the transformation
\eqn\ej{\eqalign{ M_d \rightarrow TM_dS_1^\dagger=M_d^{diag}
          \qquad  M_u \rightarrow TT'^\dagger(T'M_uS_2^\dagger)T'^{*}T^{T}=
                      V_{KM}^{\dagger} M_u^{diag} V_{KM}^{*} }}
where $(T,S_1)$ and $(T',S_2)$ diagonalise $M_d$ and $M_u$ respectively and
the KM matrix $V_{KM}=T'T^{\dagger}$.

\centerline{\bf 3. Perturbative Minimisation of Higgs Potential}

	We consider the most general renormalizable potential of the
representations $\eh$:
\eqn\emw{\eqalign{\lambda_1^1(tr(\Phi_{1\alpha}\Phi_1^{\alpha})-C_1^2)^2 +
\lambda_2^1((tr\Phi_{1\alpha}\Phi_1^\alpha)^2
&-tr(\Phi_{1\alpha}\Phi_1^\alpha)^2)\cr  +\lambda^1_3((tr\Phi_{1\alpha}
\Phi_1^{\alpha})^2-tr(\Phi_{1\alpha}\Phi_1^{\beta})
tr(\Phi_{1\beta}\Phi_1^{\alpha}))&+(1\rightarrow 2)\cr }}
\eqn\emo{\eqalign{+\lambda_1^{12}(tr(\Phi_{1\alpha}\Phi_1^\alpha)
tr(\Phi_{2\beta}\Phi_2^\beta)-tr(\Phi_{1\alpha}
\Phi_1^\beta)tr(\Phi_{2\beta}\Phi_2^\alpha))}}
\eqn\emp{\eqalign{ +\lambda_2^{12}tr(\Phi_{1\alpha}
\Phi_{2\beta})tr(\Phi_1^\alpha \Phi_2^\beta) +\lambda_3^{12}
tr(\Phi_{1\alpha}\Phi_{2\beta})tr(\Phi_1^\beta \Phi_2^\alpha) }}
\eqn\emx{\eqalign{+\lambda_4^{12}tr(\Phi_{1\alpha}\Phi_{2\beta} \Phi_2^\beta
\Phi_1^\alpha) +\lambda_5^{12}tr(\Phi_{1\alpha}\Phi_{2\beta} \Phi_2^\alpha
\Phi_1^\beta) }}
\eqn\emr{\eqalign{+\mu_1^{11}\Phi_{1ab}\Sigma^{1b}\Sigma^1_c\Phi_1^{ac} +
\mu_1^{22}\Phi_{1ab}\Sigma^{2b}\Sigma^2_c\Phi_1^{ac}
+\mu_2^{11}\Phi_{2ab}\Sigma^{1b}\Sigma^1_c\Phi_2^{ac}
+\mu_2^{22}\Phi_{2ab}\Sigma^{2b}\Sigma^2_c\Phi_2^{ac} }}
\eqn\ems{\eqalign+{\mu_1^{12}\Phi_{1ab}\Sigma^{1b}\Sigma^2_c\Phi_1^{ac} +h.c
+\mu_2^{12}\Phi_{2ab}\Sigma^{1b}\Sigma^2_c\Phi_2^{ac}  +h.c.   }}
\eqn\emv{\eqalign{ +\lambda^{ij}\Phi_{1ab}\Sigma^i_m
\Sigma^j_n\Phi_{2ef}\epsilon^{ame}\epsilon^{bnf}+h.c. }}
    The traces are over $SU(3)$ indices and the indices raised relative to
$\eh$ indicate complex conjugation. We have ignored terms like
 $\Phi_{1ab}\Sigma^{1c}\Sigma^1_c\Phi_1^{ab}$ which contribute to
the overall magnitude of the vevs but do not affect their
relative orientation.  We have also
not explicitly written the different $SU(5)$
contractions in the $\Sigma-\Phi$ terms. To simplify slightly we have
relegated the analysis of the ({\bf 24},{\bf 3})s' potential to an appendix. We
show
there that the vevs can give the desired breaking pattern
\eqn\eq{\eqalign{\Sigma^1=(v_1diag(1,1,1,-{3 \over 2},-{3 \over2}), 0,0)
 \qquad \Sigma^2=( 0,v_2diag(1,1,1,-{3 \over 2},-{3 \over2}),0)}}
When we substitute these vevs in the potential above the different $SU(5)$
contractions give the same term. In the same way as in minimal $SU(5)$ a
fine tuning between these two contractions is used to make the colour
triplets heavy.

    The problem of finding the minimum of this potential
is analytically intractable for general values of the couplings. What
we do now is show that we can produce approximate minima of
the form required to fit the phenomenology in some chosen
region of coupling parameter space.

    Consider first the self couplings of the $\Phi$ fields $\emw$.
      For positive
couplings we have a global minimum at
$\Phi_{\alpha a b} = C\delta_{\alpha}^5 \delta_a^3\delta_b^3$.
The only degeneracy on
the minimum is given by the action of $SU(5)\times SU(3)$. This can be seen
by expanding to quadratic order about the minimum and checking that the only
massless degrees of freedom are the Goldstone modes
 generated by the action of the group.
We can see qualitatively that the third term pushes the six $SU(5)$ vectors
parallel.
If we then rotate them into one direction (the ``5'' direction, say)
the second term is of the form $(trA)^2 - trA^2$
where $A=\Phi_5 \Phi^5$ is an hermitian matrix with positive
eigenvalues. To make this zero we must have
\eqn\eo{\eqalign{A=v^2{\pmatrix{0 &0 &0 \cr 0 &0 &0 \cr 0 &0 &1 \cr}}\quad
	   \Rightarrow \quad \Phi_5=C{\pmatrix{0 &0 &0 \cr 0 &0 &0 \cr 0 &0 &1 \cr}}}}

    We can next use the positive definite term $\emo$ to make the two
representations parallel in $SU(5)$ and, rotating them into the 5 direction,
we have vevs
\eqn\eoo{\eqalign{\Phi_1=C_1{\pmatrix{0 &0 &0 \cr 0 &0 &0 \cr 0 &0 &1
\cr}}\quad
	   \Phi_2=C_2U{\pmatrix{0 &0 &0 \cr 0 &0 &0 \cr 0 &0 &1 \cr}}U^T }}
where $C_1,C_2 \in I\!R$ and $U \in SU(3)$. If we now couple to the $\Sigma$
vevs in $\eq$ through the first three terms in $\emr$ we are left with only
an $SU(2)$ degeneracy on the minimum and the vevs can be written
\eqn\eo{\eqalign{\Phi_1=C_1{\pmatrix{0 &0 &0 \cr 0 &0 &0 \cr 0 &0 &1 \cr}}\quad
	    \Phi_2=C_2e^{2i\phi}{\pmatrix{0 &0 &0 \cr 0 &\beta^2e^{2i\psi} &
\alpha\beta e^{i\psi} \cr 0 &\alpha\beta e^{i\psi} &\alpha^2 \cr}}}}
where $\alpha^2+\beta^2=1$ where $\alpha,\beta \in I\!R$.
The terms $\emp$ and $\emx$ each reduce to just one when we fix the $SU(5)$
direction and they are
\eqn\eu{\eqalign{tr(\Phi_1\Phi_2)tr(\Phi^*_1\Phi^*_2)\qquad
                   tr(\Phi_1\Phi_2\Phi^*_2\Phi^*_1) }}
These together with the last term in $\emr$ produce
a quadratic in $ \beta^2$
\eqn\eua{\eqalign{(\mu^{22}_2C_2^2v_2^2 - \lambda^{12}_4C_1^2C_2^2
-2\lambda^{12}_2C_1^2C_2^2)\beta^2+\lambda^{12}_2C_1^2C_2^2\beta^4}}
which we can use to fix $\beta$.
($\lambda^{12}_2$ and $\lambda^{12}_4$ are here the sums of the appropriate
couplings).

All the other parameters we shall need to produce acceptable mass
matrices may be obtained by considering fluctuations about the
above form, and balancing linear corrections coming from terms in the
potential not so far considered
against quadratic mass terms.
We consider all
perturbations to the leading order vevs together and write them in the
form
\eqn\ewa{\eqalign{\Phi_1 \rightarrow C_1\pmatrix{0 &0 &0 \cr 0 &0 &0 \cr 0 &0
&1 \cr}
   + C_1\pmatrix{\alpha_1 &\rho_1 &\rho_2 \cr
         \rho_1 &\alpha_2 &\rho_3 \cr \rho_2 &\rho_3 &\alpha_3 \cr}}}
%
\eqn\ewb{\eqalign{\Phi_2  \rightarrow C_2e^{2i\phi}{\pmatrix{0 &0 &0 \cr 0
&\beta^2e^{2i\psi} &    \alpha\beta e^{i\psi} \cr 0 &\alpha\beta e^{i\psi}
&\alpha^2 \cr}}                    +C_2e^{2i\phi}\pmatrix{\beta_1 &\delta_1
&\delta_2 \cr
         \delta_1 &\beta_2 &0 \cr \delta_2 &0 &\beta_3 \cr}}}
where $\beta_3 \in I\!R$.

	We now calculate the linear and quadratic corrections to the potential order
by order
in the small parameter $\epsilon \sim  {\sqrt \beta}$. We take the
terms $\eu$
to be of order $\epsilon$ smaller than the leading terms which gave us $\eo$
and
the remaining terms $\ems$ and $\emv$ to
be of order $\epsilon^2$ i.e.
\eqn\emagn{\eqalign{ \lambda_i^1 C_1^4 \sim
 \lambda_i^2 C_2^4 \sim  \lambda_1^{12}C_1^2C_2^2
\sim  \mu_1^{11}C_1^2v_1^2 \sim
 \mu_1^{22}C_1^2v_2^2 \sim  \mu_2^{11}C_2^2v_1^2 \cr
 \lambda_2^{12}C_1^2C_2^2 \sim \lambda_4^{12}C_1^2C_2^2 \sim
\mu_2^{22}C_2^2v_2^2 \sim \epsilon \lambda_1^1 C_1^4 \cr
\mu_1^{12}C_1^2v_1v_2 \sim  \mu_2^{12}C_2^2v_1v_2 \sim
 \lambda^{ij}v_iv_jC_1C_2 \sim \epsilon^2 \Lambda_1^1 C_1^4 \cr }}

The leading linear corrections are
\eqn\ell{\eqalign{(\lambda^{12}_2+\lambda^{12}_4)C_1^2C_2^2(\alpha_3 +
\beta_3+h.c.)}}
which are just corrections to the entries which we are already free to
fix at leading order.
To order $\epsilon^2$ the linear corrections are
\eqn\ey{\eqalign{\lambda^{11}C_1C_2e^{-2i\phi}v_1^2(\alpha_2 + \beta_2^*) +
h.c.\cr
           -\lambda^{12}C_1C_2e^{-2i\phi}v_1v_2(\rho_1 + \delta_1^*) + h.c.\cr
           \lambda^{22}C_1C_2e^{-2i\phi}v_2^2(\alpha_1 + \beta_1^*) +h.c.\cr }}
The quadratic corrections to the leading terms are

%
\eqn\ema{\eqalign{\lambda_1^1C_1^4(\alpha_3^*+
\alpha_3)^2&+\lambda_1^2C_2^4(\beta_3^*+\beta_3)^2
\cr
+2\lambda_2^1C_1^4(|\alpha_1|^2+|\alpha_2|^2+2|
\rho_1|^2)&+2\lambda_2^2C_2^4(|\beta_1|^2+|\beta_2|^2+2|\delta_1|^2)\cr
  +\mu_1^{11}v_1^2C_1^2(|\alpha_1|^2+|\rho_1|^2+|\rho_2|^2)
&+\mu_1^{22}v_2^2C_1^2(|\rho_1|^2+|\alpha_2|^2+|\rho_3|^2)\cr 
&+\mu_2^{11}v_1^2C_2^2(|\beta_1|^2+|\delta_1|^2+|\delta_2|^2)\cr}}
Fixing the perturbations in $\ey$ with
$\ema$ we get to order $\epsilon^2$
\eqn\ewp{\eqalign{<\Phi_1>&=C_1\pmatrix{\alpha_1e^{i(2\phi-\theta_{22})}
&\rho_1e^{i(2\phi-\theta_{12})} &0\cr
         \rho_1e^{i(2\phi-\theta_{12})} &\alpha_2e^{i(2\phi-\theta_{11})} &0
\cr 0&0 &1 \cr}  \cr <\Phi_2>&=
C_2e^{2i\phi}{\pmatrix{\beta_1e^{-i(2\phi-\theta_{22})}
&\delta_1e^{-i(2\phi-\theta_{12})} &0 \cr \delta_1e^{-i(2\phi-\theta_{12})}
&\beta_2e^{-i(2\phi-\theta_{11})} & \beta e^{i\psi} \cr 0 &\beta e^{i\psi} &1
\cr}}\cr }}
where we now use $\alpha_1,\alpha_2$ etc.to denote the magnitudes of the
perturbations and
$\theta_{ij}$ are the phases in the $\lambda^{ij}$ of $\emv$.

	The phases $\psi$ and $\phi$ are still undetermined as the
dependence on them vanishes when we substitute back the perturbations
in $\ewp$ into $\ey$ and
$\ema$. To fix
them we must go to higher order in our perturbative minimization.
At next order we get the additional linear correction
\eqn\eyt{\eqalign{ (2\lambda^{12}_2+\lambda^{12}_4)C_1^2C_2^2 \beta
e^{i\psi}\rho_3 +h.c. }}
and quadratic corrections
\eqn\emb{\eqalign{\lambda^{12}_2C_1^2C_2^2((\alpha_1\beta_1 +\alpha_2
\beta_2 +\alpha_3\beta_3+\alpha_3\beta_3^*&+ 2\rho_1\delta_1+
2\rho_2\delta_2+h.c.)+ |\alpha_3|^2+|\beta_3|^2)\cr
+\lambda^{12}_4C_1^2C_2^2(|\rho_2|^2+|\delta_2|^2+|\rho_3|^2
+|\alpha_3|^2+|\beta_3|^2&+(\rho_2\delta_2+\alpha_3\beta_3
+\alpha_3\beta_3^* +h.c. ))\cr+\mu^{22}_2C_2^2v^2_2
(|\delta_1|^2&+|\beta_2|^2)\cr}}
 	If we consider all terms to this order we have for each pair
of corrections a quadratic potential of the form
\eqn\eyx{\eqalign{
\lambda\delta+\lambda'\rho^*+h.c.+\lambda_1|\delta|^2
+\lambda_2|\rho|^2+\lambda_3(\delta \rho+h.c.)  }}
which has minima at
\eqn\eym{\eqalign{\rho=-{1\over{\lambda_0}}(\lambda_1\lambda'-\lambda_3\lambda)
     \qquad \delta=-{1\over{\lambda_0}}(\lambda_2\lambda^*-\lambda_3\lambda'^*)
             }}
and terms which depend on the phases of the couplings which can be written
\eqn\eyv{\eqalign{{{\lambda_3}\over{\lambda_0}}(\lambda\lambda'^* +h.c.)}}
where $\lambda_0=\lambda_1\lambda_2-\lambda_3^2$.
In $\ey$ and $\eyt$ $\lambda\lambda'^*$ is real and we again fail to find
contributions to the potential to fix the phases.

At order $\epsilon^4$ we get the linear terms
\eqn\eyy{\eqalign{-2\lambda^{11}C_1C_2e^{-2i\phi}v_1^2 \beta e^{-i\psi}\rho_3 +
h.c.\cr
           \lambda^{12}C_1C_2e^{-2i\phi}v_1v_2 \beta
e^{-i\psi}\rho_2 + h.c.\cr       \mu^{12}_2C_2^2v_1v_2 \beta
e^{-i\psi} \delta_2 +h.c.\cr }}
and quadratic terms
\eqn\eqt{\eqalign{ -2\lambda_2^2C_2^4\beta
              e^{i\psi}\delta_2\delta_1^*&+h.c. \cr
   +\mu_1^{12}v_1v_2C_1^2(\alpha_1\rho_1^*+\rho_1\alpha_2^*+\rho_2\rho_3^*)&+
   \mu_2^{12}v_1v_2C_2^2(\beta_1\delta_1^*+\delta_1\beta_2^*) +h.c. \cr
\lambda^{11}C_1C_2e^{-2i\phi}v_1^2(
+\alpha_3\beta_2^*&+\alpha_2\beta_3^*)+h.c.\cr
           +\lambda^{12}C_1C_2e^{-2i\phi}v_1v_2(
+\rho_3\delta_2^*&-\alpha_3\delta_1^* -\rho_1\beta_3^*)+h.c.\cr
  \lambda^{22}C_1C_2e^{-2i\phi}v_1^2 (-2\rho_2\delta_2^* &+\alpha_1\beta_3^*
+\alpha_3\beta_1^* ) + h.c.\cr }}
When we substitute the corrections fixed at higher orders into these
terms the dominant contribution which depends on the appropriate
phases is
\eqn\edc{\eqalign{4
\beta^2
C_1C_2^3({{v_1}\over{v_2}})^2{{\lambda^{11}(2\lambda_2^{12}+\lambda_4^{12})}
                    \over{\mu_1^{22}} }    cos(2\psi+2\phi-\theta_{11}) }}
The other terms which depend on these phases are smaller by
$\epsilon$ or more. Thus to this order the argument of the cosine is
fixed to be $0$ or $\pi$ at the minimum depending on the sign
of the coefficient.

	We find however that we only get terms which fix $\psi+\phi$.
The reason for this is accidental $U(1)$
symmetries in the low energy tree level Lagrangian. Without the terms
$\emv$ these are
\eqn\eas{\eqalign{\Phi_1 \rightarrow e^{i\alpha}U_\theta \Phi_1 U_\theta^T
\quad
                  \Phi_2 \rightarrow e^{i\beta}e^{-2i\theta} U_\theta^* \Phi_1
U_\theta^{\dagger} \quad \Psi \rightarrow e^{i(\alpha +{\beta \over
2})}e^{-i\theta} U_\theta \Psi
                   \quad \chi \rightarrow e^{-i{\beta \over 2}}e^{i\theta}
U_\theta \chi }}
where $\Psi$ and $\chi$ are the fermion ${\bf 3}$s and $U_\theta$ is the
diagonal family transformation
\eqn\ewt{\eqalign{U_\theta=\pmatrix{e^{i\theta} &0 &0\cr
         0 &e^{i\theta} &0 \cr 0&0 &e^{-2i\theta} \cr} }}
The terms $\emv$ break $\eas$ to a remaining two $U(1)$s by forcing
$\alpha=\beta$. The first $U(1)$ which is an accidental symmetry of
the whole theory is then just the usual one which
becomes B-L when combined with the appropriate $SU(5)$ generator.
The second ``$\theta$'' symmetry will presumably be broken by higher order
quantum corrections.

	We do not need to consider these contributions however as the
action on the fermions in $\eas$ is just a quark phase
redefinition and can be used to transform the mass matrices into the form
\eqn\emm{\eqalign{M_d&=m_b\pmatrix{\alpha_1e^{i(2\phi+2\psi-\theta_{22})}
&\rho_1e^{i(2\phi+2\psi-\theta_{12})} &0\cr
         \rho_1e^{i(2\phi+2\psi-\theta_{12})}
        & \alpha_2 e^{i(2\phi+2\psi-\theta_{11})} &\rho_3 \cr 0&\rho_3 &1 \cr}
\cr M_u&=m_t\pmatrix{\beta_1e^{i(2\phi+2\psi-\theta_{22})}
&\delta_1e^{i(2\phi+2\psi-\theta_{12})} &0 \cr
\delta_1e^{i(2\phi+2\psi-\theta_{12})} &\beta_2e^{i(2\phi+2\psi-\theta_{11})} &
\beta \cr 0 &\beta &1 \cr}\cr }}
(to order $\epsilon^3$) where $\rho_3$ is now also real. Note that
$M_u \propto \Phi_2^*$. A further quark phase redefinition yields
\eqn\emmb{\eqalign{M_d=m_b\pmatrix{\alpha_1e^{i\theta'} &\rho_1 &0\cr
         \rho_1
        & \alpha_2 e^{i\phi'} &\rho_3 \cr 0&\rho_3 &1 \cr}
M_u=m_t\pmatrix{\beta_1e^{i\theta'} &\delta_1 &0 \cr \delta_1
&\beta_2e^{i\phi'} & \beta \cr 0 &\beta &1 \cr} }}
where $\theta'=-(2\phi+2\psi-2\theta_{12}+\theta_{22})$ and
$\phi'=2\phi+2\psi-\theta_{11}$.

	The leading terms which fix $\beta^2$ are just those in $\eua$ and we
thus have
\eqn\eub{\eqalign{\beta^2={{(2\lambda^{12}_2C_1^2C_2^2-\mu^{22}_2C_2^2v_2^2 +
\lambda^{12}_4C_1^2C_2^2
)}\over{2\lambda^{12}_2C_1^2C_2^2}} }}
and we require
$0<\mu^{22}_2C_2^2v_2^2 - \lambda^{12}_4C_1^2C_2^2<2\lambda^{12}_2C_1^2C_2^2$
to make this a minimum. Using $\edc$ we can now see that
the 22 corrections (which are proportional to
$\lambda^{11}$) in both
mass matrices in $\emmb$ are determined to be real quantities which are
positive or negative depending on whether the sign of
$2\lambda^{12}_2+\lambda^{12}_4$ is positive
or negative. The sign of $\rho_3$ is then determined by $\eyt$ to be
opposite to that of $\alpha_2$.

	Choosing the negative sign for $\alpha_2$ we arrive at the form
\eqn\emf{\eqalign{M_d=m_b\pmatrix{\alpha_1e^{i\theta '} &\rho_1 &0\cr
         \rho_1
&-\alpha_2 &\rho_3 \cr 0&\rho_3 &1 \cr}  \qquad
M_u=m_t\pmatrix{\beta_1e^{i\theta'}
& \delta_1 &0 \cr \delta_1 &-\beta_2& \beta \cr 0 &\beta &1 \cr}   }}

	Although there are ten real parameters and one phase this
is a constrained form of the mass matrices.
We will outline the procedure below which can be used to diagonalize
these matrices and extract the phenomenological predictions. We will
not consider this particular form in detail however as we need to
add further Higgs fields to this model to make it realistic.
We will only note that a full calculation shows the CP violation in
these mass matrices to be very small as the only complex elements
are forced very small by the hierarchy in the quark masses.

	One comment on the perturbative method used to generate these
mass matrices is required. We have shown that we can generate from
the potential
under the given assumptions mass matrices with entries at certain
orders in the expansion parameter $\epsilon$
\eqn\emoo{\eqalign{M_d\sim\pmatrix{\epsilon^2 &\epsilon^2 &\epsilon^4 \cr
                                   \epsilon^2 &\epsilon^2 &\epsilon^3
\cr                                \epsilon^4 &\epsilon^3 &1 \cr} \qquad
                  M_u \sim \pmatrix{\epsilon^2 &\epsilon^2 &\epsilon^4 \cr
                                    \epsilon^2 &\epsilon^2 &\epsilon^2
\cr                                 \epsilon^4&\epsilon^2 &1 \cr}  }}
This perturbation
expansion in $\epsilon$ can however be altered easily to give the
corrections at the orders needed to fit phenomenological mass matrices.
For example if we assume
\eqn\esa{\eqalign{ ({ {v_2}\over{v_1}})^2 \sim \epsilon \quad
                    { {C_1}\over{C_2}} \sim \epsilon \quad
 \lambda_1^2 \sim \lambda_2^2 \sim \lambda_3^2 \sim \lambda_2^{12}
\sim \lambda_4^{12}\quad   \cr
                   \mu_1^{11}C_1^2v_1^2 \sim \lambda_2^2C_2^4 \quad
                   \lambda^{11}\sim\lambda^{12}\sim\lambda^{22}\quad
                   \mu_1^{11}\sim\mu_2^{11}\sim\mu_1^{22}\quad \cr
                   \mu_2^{22}\sim \epsilon^2 \mu_1^{11}\quad
                   \mu_1^{12}\sim \mu_2^{12}\sim \epsilon^6 \mu_1^{11}\quad
                   \lambda^{11}\sim \epsilon^4 \mu_1^{11} \cr }}
we get
\eqn\emop{\eqalign{M_d\sim \pmatrix{\epsilon^4 &\epsilon^3 &\epsilon^5 \cr
                                    \epsilon^3 &\epsilon^2 &\epsilon^3
\cr                                 \epsilon^5 &\epsilon^3 &1 \cr} \qquad
                  M_u \sim \pmatrix {\epsilon^6 &\epsilon^5 &\epsilon^7 \cr
                                     \epsilon^5 &\epsilon^3
&\epsilon^2 \cr                      \epsilon^7 &\epsilon^2 &1 \cr}  }}
It is not difficult to see that the phase analysis is also unchanged with
these assumptions.

\centerline{\bf 4.  FCNCs and Higgs Phenomenology}

 Tree level flavour changing neutral currents (FCNCs)
  arise generically in models of this type. From the Yukawa
couplings e.g.
\eqn\efc{\eqalign{ Y \psi_a^{\alpha T}C\chi_{b \alpha \beta }\Phi_1^{\beta
ab}}}%
we see that they are mediated by the off diagonal elements of
the ${\bf 6}$ and $\overline{{\bf 6}}$. When we include the effects of quark
mixing
there will also be an induced mixing in these currents.

    The obvious way of suppressing such processes below
their phenomenological
limits is by appropriately raising the masses of the offending particles.
In a single Higgs doublet model the neutral Higgs after SSB has mass $m$
with $m^2\approx \lambda v^2$ where $v$ is the vev, fixed by the
gauge boson mass. To make $m$
large we must move into the strong coupling region which is unattractive
and may violate perturbative unitarity bounds. However we see that in $\ema$
there are
many other terms which contribute to the Higgs masses. In particular the
$\mu$ terms provide mass contributions which are not related to a
4 point low energy coupling. In fact precisely the sort of choices we
made in $\esa$ to produce the elements in the mass matrices at the
appropriate order are those needed to suppress the FCNCs. For example
$ \mu_1^{11}C_1^2v_1^2 \sim \lambda_2^2C_2^4$ implies
$ \mu_1^{11}v_1^2 \sim \epsilon^{-2}\lambda_2^2C_2^2$.
 $\mu_1^{11}v_1^2$ is precisely the mass squared picked up
through this term in the potential by the
perturbations mediating the $d\rightarrow q$ currents and
the term on the right side of the relation
is the mass squared associated with the four point coupling $\lambda_2^2$.

     The only
components which do not pick up a mass in this way are the Higgs
predominantly mediating the flavour neutral processes in the b and t
quarks. Even with  mixing they only mediate the u$\rightarrow$t
and c$\rightarrow$t. Thus the suppression of FCNCs and the existence of one
heavy quark of each charge are tied together. The same terms in the
potential which make the FCNC mediating Higgs heavy single out the direction
of the heavy family and suppress the masses of the other quarks. The
heaviest Higgs particles mediate processes between the lightest
quarks.

    In the particular model we discussed above this feature is frustrated
by the requirements of generating $\beta$. We chose to put this parameter
where we did because the Higgs particle which gets mass from $\mu^{22}_2$
is the c$\rightarrow$t mediating one. We required
$\mu^{22}_2 <\lambda^{12}_4+2\lambda^{12}_2$ (dropping vevs).
This had the effect of making the dominant contribution to the mass of
the $\beta_2$ fluctuation come from the $\lambda_2^2$ term. Then the
requirement that $\alpha_2$ and $\beta_2$ be generated at $\epsilon^2$
and $\epsilon^3$ respectively forced us to choose
$\mu_1^{22}C_1^2v_2^2\sim \epsilon \lambda_2^2C_2^4$ which leads to
$\mu_1^{22}v_2^2\sim \epsilon ^{-1}\lambda_2^2C_2^2$. The $\alpha_2$
mediates significant $d\rightarrow s$ when we take into account the
quark mixing in the mass eigenstate basis and we will again be forced
to making the four point couplings large to suppress them to the
desired level. In the
extensions of this model which we discuss
below this is not the case.

    One caveat is required here. As noted in the introduction these terms
which may give extra mass to the Higgs particles are also crucial in the
generation of quantum corrections. It is not difficult to see that the
radiative corrections become important if $\mu v^2 \sim \lambda C^2$. If one
allows these cross terms to become large we would have to address the
question in detail of the corrections to this tree level analysis and
what sort of additional fine-tuning might be required. We shall not
do this here - we merely note that there are terms present in the
Lagrangian which may be used to give masses to the FCNC Higgs perhaps
an order of magnitude larger than the electroweak scale. A more
detailed analysis would be required to determine to what extent
fine tuning is necessary to preserve this mild heirarchy of mass scales.

    It is worth emphasising that the low energy Higgs sector of this theory
would have a rich and distinctive phenomenology. Up to mixing
each of the neutral components of the twelve Higgs
doublets mediates a different neutral current. The gauge boson
couplings are also interesting - at leading order only the doublets
mediating $t\overline{t}$ and $b\overline{b}$ processes have linear
couplings to the gauge bosons.

\centerline{\bf 5. Radiative Corrections and Running }

    The mass matrices we extract from the potential are valid at the GUT
scale. We implicitly assumed a simple scaling of these matrices to 1GeV.
This is in fact exact. To see this consider the one loop  1PI
corrections to the Yukawa couplings, masses
and mixing angles involving one internal Higgs line. The Yukawa couplings of
the neutral and charged
Higgs to the quarks are
%
\eqn\eay{\eqalign{Y{\overline{q'}}_{aR}q'_{bL}\Phi_1^{5ab}+
Y^*{\overline{q'}}_{aL}q'_{bR}\Phi_{15ab}&+
Y'{\overline{q}}_{aR}q_{bL}\Phi_{25}^{ab}+
Y'^*{\overline{q}}_{aL}q_{bR}{{\Phi_2}^5}_ab\cr
+Y{\overline{q'}}_{aR}q_{bL}\Phi_1^{4ab}
+Y^*{\overline{q}}_{aL}q'_{bR}\Phi_{14ab}
&-Y'{\overline{q}}_{aR}q'_{bL}\Phi_{24}^{ab}
-Y'^*{\overline{q'}}_{aL}q_{bR}{{\Phi_2}^4}_ab \cr}}%
where $q$ and $q'$ are the u and d type quarks respectively.
It is easy to see from these couplings that in the unbroken phase these
diagrams are not allowed
because they do not conserve global charge. If they are non
zero in the broken phase they must derive from a term in the unbroken
effective action with more $\Phi$ or $\Sigma$ external legs. Any such
term is finite however just because of renormalizability. (The mass
self energy with one extra leg is just the first term). Therefore the
mass matrices do not run at all due to Higgs couplings. All the running
comes from the gauge fields which distinguish only between the three mass
matrices but not their  elements.

      Running and finite radiative corrections are particularly relevant
if one extracts a predictive ansatz by imposing symmetries on the Higgs
potential. Since the running just causes a simple scaling in these
models we must turn to finite corrections to generate anything we do
not have at tree level. We have considered such corrections and found them
to be far too small to serve this purpose. To generate a
correction $\approx{1\over {48\pi^2}} \approx 10^{-3}$ we find that
we need precisely the term
in the potential which would generate the parameter at tree level. Any other
potentially interesting corrections are generated only at $\approx{1\over
{48\pi^2}}\beta$. One noteworthy feature which could perhaps be exploited
usefully in a different model is that the corrections contribute
differently to the $M_d$ and $M_e$ matrices (e.g. if the tree level vev
giving the radiative correction did not couple to the fermions because
of some symmetry).

\centerline{\bf 6. Alterations to the minimal $SU(5)$ model }

    The model we discussed above has several shortcomings:

(i) We ignored the lepton mass matrices. In $SU(5)$ the Yukawa couplings
produce the mass relation $M_d=M_e$ at the GUT scale. When renormalized
to the electroweak scale this gives
the successful relation $ m_b \approx 3m_\tau$ \ref\massrun{A.J.Buras et al.,
Nuc. Phys. {\bf B135} (1978) 66.}. However the other relations
are not right. Georgi and Jarlskog
\ref\gj{G.Georgi and C.Jarlskog,Phys.
Lett. {\bf 86B} (1982) 297.}
proposed
alternative mass relations between the charge -${1\over3}$ quarks and
the leptons which they showed could be produced with the addition of a
Higgs {\bf 45} of $SU(5)$. We thus add a ({\bf 45},{\bf 6})
which couples to the first term
in $\ea$ only.

(ii) The model does not unify as it is at low energy just the standard
model with extra doublets
\ref\unif{A. Giveon, L. Hall and U. Sarid, Phys. Lett. {\bf 271B} (1991) 138.}.
We could consider many types of additions of heavier particles to remedy
this as in \ref\add{U. Amaldi, W. de Boer, P. Frampton, H. Furstenau, J. Liu,
Phys. Lett. {\bf 281B} (1992) 374.}.
However the addition of the
$({\bf 45},{\bf 6})$ prompted by (i) is already adequate for this purpose.

        In \unif\  it was shown that the addition of a single ${\bf 45}$ to
the one Higgs doublet minimal $SU(5)$ model is sufficient to bring about
unification if some of the components of this ${\bf 45}$ are put at
intermediate mass scales e.g. if the $(3,3)$
($SU(3)_c \times SU(2)_w$) component is at $10^8-10^9$ GeV and the $(8,2)$
component at some scale below this. One must calculate
beta functions with contributions from an ``effective number'' $n_{eff}$
of copies of each multiplet, given by
\eqn\eum{\eqalign{
n_{eff}=\Sigma_i n_i{{log {M_G \over M_i}\over log{M_G \over M_Z}}}  }}
where $n_i$ are the number of the appropriate multiplet at
mass  $M_i$, $M_G$ is the unification scale
(the X boson mass) and $M_Z$ is the Z mass.
With the eighteen electroweak doublets
and six ${\bf 45}$s in our model we unify with components at intermediate
mass scales closer to the GUT scale. We have found, for example, that with
the $(\overline{6},1)$ component at $10^{10}$GeV and the $(8,2)$ component
at $10^{13}$GeV (for all six copies) we have $M_G$ at $10^{15}$GeV. For a
higher unification scale of $4 \times 10^{16}$ we put both components
at
 $10^{12}$GeV
\ref\unidata{In the notation of ref. \unif\  the $(8,2)$ contributes
$-{8\over15}$ to $B_{12}$ and $-{2\over3}$ to $B_{23}$, and
the $(\overline{6},1)$ contributes
$+{2\over15}$ to $B_{12}$ and $-{5\over6}$ to $B_{23}$. These figures are
quoted in the preprint (LBL-31084, UCB-PTH 91/35) of \unif\ but excluded
from the published version.}. We could also investigate the
possibility of using some of the components of the $({\bf 24},{\bf 3})$
to alter the running. The main point here is that
because there are many copies of the $SU(5)$
representations  significant contributions to the beta functions can come
from fields not far below the GUT scale.

(iii) In the analysis of the simplest model above we noted that
 we must make Higgs self-couplings large to
suppress FCNCs. With an extra {\bf 45} however we expect the $\alpha_2$
correction to come predominantly from the {\bf 45}, to generate the G-J mass
relations. Then there is no problem if we make $\mu^{22}_1$ larger. In
general the smallness of a correction to the leading order form is
associated with a large Higgs mass for the corresponding degree of
freedom.

         We will only briefly discuss the effects of adding
a $({\bf 45},{\bf 6})$ to the Higgs potential as the analysis is just
as in the case we have looked at in detail.

	A ${\bf 45}$ of $SU(5)$ can be represented as a three
index tensor ${H_{\alpha\beta}}^\gamma$ with the properties
\eqn\effa{\eqalign{{H_{\alpha\beta}}^\gamma=-{H_{\beta \alpha}}^\gamma
                 \qquad {H_{\alpha\beta}}^\alpha=0 }}
It breaks $ SU(5)$ $\downarrow$ $SU(3) \times SU(2) \times U(1)$
with the vev
\eqn\effb{\eqalign{<{H_{\alpha\beta}}^\gamma>&\propto \delta_\beta^5
 (\delta_\alpha^\gamma-4\delta_4^\gamma \delta_\alpha^4) -(\beta
\leftrightarrow
\alpha)}}
The $({\bf 45},{\bf 6})$ couples to the first term in $\ea$ only and
for a contribution $m$ to an element of the $M_d$ mass matrix gives $-3m$
for the corresponding element in the lepton mass matrix.

	 We can construct a Higgs potential for the
$({\bf 45},{\bf 6})$ alone which has zero vev. Coupling to the
$({\bf5},{\bf 6})$ and $({\bf 5},{\bf \overline{6}})$ we find that the
$({\bf 45},{\bf 6})$ is generated in the appropriate $ SU(5)$
directions in $\effb$. Then considering the $SU(3)$
perturbations we find that the only corrections generated are
\eqn\effc{\eqalign{\hat{\lambda}^{11}C_1C_2e^{-2i\phi}v_1^2\hat{\alpha}_2 +
h.c.\cr
           -\hat{\lambda}^{12}C_1C_2e^{-2i\phi}v_1v_2\hat{\rho}_1  + h.c.\cr
           \hat{\lambda}^{22}C_1C_2e^{-2i\phi}v_2^2\hat{\alpha}_1 +h.c.\cr }}
and at $\epsilon^2$ smaller
\eqn\effd{\eqalign{-2\hat{\lambda}^{11}C_1C_2e^{-2i\phi}v_1^2 \beta
e^{-i\psi} \hat{\rho}_3 + h.c.\cr
          \hat{\lambda}^{12}C_1C_2e^{-2i\phi}v_1v_2 \beta
e^{-i\psi}\hat{\rho}_2 + h.c.\cr }}
where the hats are used everywhere to denote the replacement of the
$({\bf 5},{\bf 6})$ by the $({\bf 45},{\bf 6})$ (which has the same
$SU(5)$ quintality and couples in the same ways to the other
fields). The  $({\bf 45},{\bf 6})$ is taken to be zero to zeroth order
in $\epsilon$ - the perturbations are written
with the same overall scale  $C_1$ to be simply compared with the $({\bf
5,6})$ perturbations..

	The only quadratic corrections involving cross couplings
between the electroweak Higgs fields are
\eqn\effe{\eqalign{
\hat{\mu}_1^{11}v_1^2C_1^2(\alpha_1\hat{\alpha}_1^*+\rho_1\hat{\rho}_1^*
+\rho_2\hat{\rho}_2^*) +h.c.\cr
\hat{\mu}_1^{22}v_2^2C_1^2(\rho_1\hat{\rho}_1^*+\alpha_2\hat{\alpha}_2^*
+\rho_3\hat{\rho}_3^*) +h.c.\cr
\hat{\mu}_1^{12}v_1v_2C_1^2(\alpha_1\hat{\rho}_1^*+\rho_1\hat{\alpha}_2^*
+\rho_2\hat{\rho}_3^*) \cr  }}
The other terms which we might expect to see by analogy with the
previous analysis turn out not to arise because of the $ SU(5)$
contractions. The analysis of the phases still holds as given because
these new terms do not give new relevant contributions.

	Thus we see that we can easily produce a G-J type ansatz
\eqn\efff{\eqalign{ M_u=m_t \pmatrix{0 & \delta_1 &0 \cr
                               \delta_1 &-\beta_2& \beta \cr
                                     0 &\beta &1 \cr}
\quad              M_d=m_b \pmatrix{0 &\rho_1 &0\cr
                       \rho_1 & \hat{\alpha}_2e^{i\theta} & \rho_3 \cr
                            0 & \rho_3 &1 \cr}
\quad              M_l=m_b \pmatrix{ 0 &\rho_1 &0  \cr
                \rho_1 & -3\hat{\alpha}_2 &\rho_3 \cr
                                   0&\rho_3 &1 \cr}  }}
simply by making $\lambda^{22}$ appropriately small and the masses of the
$\hat{\alpha}_1$ and $\hat{\rho}_1$ perturbations very large to
suppress these corrections to the mass matrix.

\centerline{\bf 7. Analysis of Masses and Mixing Angles }

	Symmetric mass matrices may be diagonalized with a single
unitary matrix
\eqn\eda{\eqalign{M_d^{diag}=V_L M_d V_L^T \qquad
         M_u^{diag}=U_L M_u U_L^T \qquad V_{KM}=U_L V_L^{\dagger} }}
To diagonalize $\efff$ and fit it to the phenomenology we follow
the procedure used in
\ref\he{X. He and W. Hou, Phys. Rev. {\bf  D41} (1990) 1517.}
and
\ref\dimo{ S. Dimopoulos, L. Hall,
and S. Raby, Phys. Rev. Lett. {\bf 68} (1992)
1984; S. Dimopoulos, L. Hall and S.
Raby,  Phys. Rev. {\bf D45} (1992) 4192.}.
 Assuming that
$\rho_3 \sim \rho_1 \sim \epsilon ^3$, $\alpha_2  \sim \epsilon^2$,
$\beta_2 \sim \epsilon^3$, $\delta_1 \sim \epsilon^5$
we can diagonalize each mass matrix
approximately with a product of two $SU(2)$ matrices
\eqn\emoq{\eqalign{V_L&= \pmatrix{ c_1 & s_1e^{i\xi_1} &0 \cr
                                 -s_1e^{-i\xi_1}  &c_1 &0
\cr                                0&0        &1 \cr}
                         \pmatrix {1 &0 &0 \cr
                                  0 &c_4 & s_4e^{i\xi_4} \cr
                                  0 &-s_4e^{-i\xi_4} & c_4\cr} \cr
                   U_L &=\pmatrix{ c_2 & s_2 &0 \cr
                                 -s_2  &c_2 &0
\cr                                0&0        &1 \cr}
                         \pmatrix {1 &0 &0 \cr
                                  0 &c_3 & s_3 \cr
                                  0 &-s_3 & c_3\cr}  \cr              }}
where the $c_1,s_1$ etc. are sines and cosines which will be given below.
Using these we get the KM matrix
\eqn\ekma{\eqalign{V_{KM}= \pmatrix{ c_1c_2-s_1s_2fe^{-i\xi_1} &
s_1c_2+c_1s_2fe^{-i\xi_1} & s_2ge^{-i\xi_1} \cr
                                 s_1c_2f^*-c_1s_2e^{i\xi_1} &
c_1c_2f-s_1s_2e^{-i\xi_1} &c_1g
\cr                                s_1g & -c_1g^* & f \cr}  }}
where $f=c_3c_4+s_3s_4e^{i\xi_4}$, $g=s_3c_4-c_3s_4e^{i\xi_4}$.
We have multiplied the first column
by $e^{i\xi_1}$ and
the first row by $e^{-i\xi_1}$ (quark phase redefinitions)
to get to this form.

	The diagonalization of the $2 \times 2$ matrices yields
\eqn\ekmb{\eqalign{
s_4\sim -\rho_3 \quad \xi_4 \sim \hat{\alpha}_2 sin \theta \sim 0 \quad
s_1 \sim {\sqrt{{m_d} \over {m_s}}} \quad  \xi_1 \sim -\theta \quad
s_3\sim \beta \quad s_2 \sim {\sqrt{{m_u} \over {m_c}}} \cr
m_s \sim
\hat{\alpha}_2 \quad m_d \sim {\rho_1^2 \over \hat{\alpha}_2} \quad
m_c\sim \beta^2 + \beta_2 \quad m_u \sim {\delta_1^2 \over m_c} \cr
m_{\tau}=m_b \quad m_{\mu} \sim
3\hat {\alpha}_2 \sim 3 m_s \quad m_e \sim {\rho_1^2 \over 3\hat{\alpha}_2}
\sim {1\over 3} m_d\cr }}
where the last three are the GUT scale G-J mass relations
which give the correct
lepton-quark mass relations when renormalized to the electroweak
scale (multiplication of the quark masses by a factor of approximately 3).

The KM matrix in the Wolfenstein parametrization is
\eqn\ekmc{\eqalign{V_{KM}= \pmatrix{ 1-{1\over2}\lambda^2 & \lambda &
A\lambda^3(\rho-i\eta) \cr
                                 -\lambda & 1-{1\over2}\lambda^2 &A\lambda^2
\cr               A\lambda^3(1-\rho-i\eta) & A\lambda^2 & 1 \cr}  }}
and we make the identifications as in \dimo
\eqn\ekmd{\eqalign{\lambda=(s_1^2+s_2^2+2s_1s_2cos \theta)^{1\over 2} \cr
                   \lambda^2 A =s_3 -s_4 \quad \lambda {\sqrt{\rho^2
+\eta^2}} =s_2 \quad \eta={s_2sin \theta \over \lambda} \cr }}
(In \dimo\ the elements $\rho_3$ and $\beta_2$ are generated as
radiative corrections to a GUT scale ansatz which has them set to
zero).
These relations are consistent with the present phenomenology
of quark and lepton masses
and the KM matrix
\ref\dataOB{At 1 GeV take $m_u$=5MeV, $m_c$=1.35GeV, $m_t$=250GeV,
 $m_d$=9MeV, $m_s$=175MeV, $m_b$=5.6GeV, $|V_{ub}|$=0.004,$|V_{cb}|$=0.044,
 $\lambda$=0.221. These give the predictions
$\sqrt{\eta^2+\rho^2}=0.28$ consistent with
$\sqrt{\eta^2+\rho^2}=0.46 \pm 0.23$,
 and $\eta$=0.27 which is also consistent \dimo.
 The 1 GeV value of $m_t$ corresponds to a
physical top mass of approximately 150GeV according to the procedure
outlined in \he. The KM matrix values are chosen from Particle Data Group,
Phys. Lett. {\bf 239B} (1990) 1, the masses (except $m_t$) from
J. Gasser and H. Leutwyler, Phys. Rep. {\bf 87} (1982) 77.} .

	Our derived form \efff\ has seven real parameters
and one phase ($\rho_3$ is small and makes little contribution
to the observable parameters)  and fits
the nine masses, three mixing angles and one phase
of the observed fermion mass matrices. There are
the usual three G-J predictions for the lepton masses and two extra
ones arising from the relations which follow from $\ekmb$ and $\ekmd$
between the quark masses and KM parameters which fix $\rho$ and $\eta$
once the other parameters are specified.

\centerline{\bf 8. Embedding in $SO(10)$ }

    The simplest possible embedding of the model we have been considering
is
\eqn\edf{\eqalign{({\bf 16},{\bf 3})&=({\bf 1},{\bf 3})+({\bf 10},{\bf 3})
+(\overline{\bf 5},{\bf 3})\cr
 ({\bf 10},{\bf 6})&=({\bf 5},{\bf 6})+(\overline{\bf 5},{\bf 6})    \cr
 ({\bf 126},{\bf 6})&=({\bf 1},{\bf 6})+(\overline{\bf 5},{\bf 6})+({\bf
45},{\bf 6})
 +...\cr
  ({\bf 45},{\bf 3})&=({\bf 1},{\bf 3})+({\bf 24},{\bf 3})+.... \cr}}
This is also a minimal choice. We must take the {\bf 126} to accommodate
the right handed neutrinos and it must
be in a {\bf 6} of $SU(3)$ to couple to the fermions. We cannot however
take the {\bf 5}s in the {\bf 10} to be the only ones to
acquire electoweak vevs as they are then forced to have the same vevs
or one to be zero by the self potential - this would not be consistent
with our perturbative scheme.
The only other components available which can
acquire vevs are precisely the ones we need - the $(\overline{\bf 5},{\bf 6})$
to
make $m_t$ large
and the $({\bf 45},{\bf 6})$ to give the G-J relations.

The restrictions
on our $SU(5)$ analysis which result are surprisingly weak. This is
simply because we have more electroweak fields so that any restriction
arising from the larger symmetry group is more than compensated for.
Again we can fit any mass matrices at tree level.
The parameter $\beta$ could be
generated between the $({\bf \overline{5}},{\bf 6})$ and GUT scale $(1,{\bf
6})$
so circumventing any
constraints on the low energy four point couplings.

    It is interesting however to speculate in this model
about one striking feature of mass matrix phenomenology - the smallness
of the $\alpha_1,\beta_1$ components. We showed above that it was
possible  to fit
the low energy phenomenology with them set to zero.
In this $SO(10)$ model the two $({\bf 24},{\bf 3})$s
are replaced by a $({\bf 1},{\bf 3})$ and a $({\bf 24},{\bf 3})$.
 One of the three couplings
of the type $\emv$ between the $({\bf 45},{\bf 6})$ and $(\overline{\bf 5},{\bf
6})$ is now
not allowed. If the $({\bf 1},{\bf 3})$ lies in the 2 direction these terms
will generate contributions to the 12 and 22 components in the
mass matrices but none to the 11 component. If we suppose that
the $({\bf 45},{\bf 6})$ and the
$({\bf 5},\overline{{\bf 6}})$ give the dominant contributions
to the corrections through these terms we get
mass matrices of the form
\eqn\ewm{\eqalign{M_u=m_t\pmatrix{0&\delta_1 &0 \cr
         \delta_1 &\beta_2 &\delta_3 \cr 0 &\delta_3 &1\cr}\quad
      M_d=m_b\pmatrix{0 &\rho_1 &0 \cr
         \rho_1 &\alpha_2 &0 \cr 0& 0&1 \cr}\quad
               M_e=m_\tau\pmatrix{0 &-3\rho_1 &0 \cr
         -3\rho_1 &-3\alpha_2 &0 \cr 0& 0&1 \cr} }}
which is successful in all but the lightest lepton-quark mass relation.

\centerline{\bf 9. Implications for Cosmic Texture}

    The investigations of cosmological structure formation and
    microwave anisotropies produced by global symmetry breaking have
    so far been performed only for the case of an $O(N)$ symmetry broken
    by a scalar ${\bf N}$ to $O(N-1)$. The case of $N=4$ is easily seen to
    be equivalent to the complete breaking of an $SU(2)$ global symmetry by
    a scalar {\bf 2}\BR,\PST. However, as we have argued here, as far as
    family symmetry goes
     a global $SU(3)$ symmetry
    appears a better bet. In this section we shall give the evolution
    equations for $SU(3)$ cosmic texture and discuss how the
    GUT scale calculated above is directly related to the parameter
    governing the amplitude of cosmological perturbations.

   The nonlinear sigma model governing the dynamics of the 8
   $SU(3)$
   Goldstone bosons is easily described in terms of the two
   complex triplets $\Sigma^1$ and $\Sigma^2$ discussed above.
   For the appropriate range of parameters in the Higgs potential,
   the minimum of the potential is given by
\eqn\ecsa{\eqalign{
\Sigma^1 &= v_1(\psi_1 +i\psi_2, \psi_3+i\psi_4, \psi_5+i\psi_6) \cr
\Sigma^2 &= v_2(\chi_1 +i\chi_2, \chi_3+i\chi_4, \chi_5+i\chi_6) \cr
\vec{\chi}^2 &= \vec{\psi}^2 =1 \qquad \qquad \vec{\psi}^T \vec{\chi}
= \vec{\psi}^T {\bf M}  \vec{\chi} =0
}}
where we suppress the matrix diag($1,1,1,-{3\over 2}, -{3\over 2}$)
in $SU(5)$ space, and the $6\times 6$
matrix ${\bf M}=$ diag $(i\sigma_2, i\sigma_2,
i\sigma_2)$ with $\sigma_2$ the usual Pauli matrix.

Following the usual procedure of imposing the constraints with
 Lagrange multipliers, one finds the following equations of
 motion for the unit vectors $\chi$ and $\psi$:
\eqn\ecsb{\eqalign{
\nabla^\mu \partial_\mu \vec{\psi} +(\partial \vec{\psi})^2 \vec{\psi}
+(\partial \vec{\psi} \partial \vec{\chi}) \vec{\chi} +
(\partial \vec{\psi} {\bf M} \partial \vec{\chi}) {\bf M} \vec{\chi} &=0 \cr
\nabla^\mu \partial_\mu \vec{\chi} +(\partial \vec{\chi})^2 \vec{\chi}
+(\partial \vec{\psi} \partial \vec{\chi}) \vec{\psi} +
(\partial \vec{\chi} {\bf M} \partial \vec{\psi}) {\bf M} \vec{\psi} &=0 \cr
}}
It is easy to see that the simplest $SU(2)$ scaling solution also
solves the $SU(3)$ equations: if one of $\vec{\psi}$ or $\vec{\chi}$
is constant it may be rotated into
  $(0,0,0,0,0,1)$. The other can then only have the first
four  components nonzero, and the equations \ecsb\ reduce to the
$SU(2)$ sigma model equation with known scaling solution
\PST. It would be interesting
to examine more general scaling solutions corresponding to other
embeddings of $SU(2)$ in $SU(3)$. It will also be interesting to perform
full three dimensional simulations of the ordering dynamics of \ecsb.

Note that the evolution equations \ecsb\
have the attractive feature that there are
{\it no} free parameters - the equations are purely geometrical,
just as in the simpler $O(N)$ theories. (It is not hard to see that
had we chosen a range of parameters in the potential so that
the minimum was at a nonzero
value of $\Sigma^{1\dagger } \Sigma^2$ there would be an extra  free
dimensionless `angle'  parameter in the evolution equations.)
However the stress energy tensor of these fields, which
determines the final density fluctuations,
does depend on the value of $v_1$ and $v_2$
separately.

Finally let us note that the relation between the unification scale
$M_X $ and  $v_1$ and $v_2$ may be used to predict the magnitude of the
cosmological density perturbations. From the standard relations (see e.g.
\ref\Ross{G.G. Ross, {\it Grand Unified Theories}, Benjamin/Cummings 1984.})
we find $M_X^2 = {25 \over 8} g^2 (v_1^2+ v_2^2)$. Substituting
the fields \ecsa\ into the scalar kinetic terms and performing
the trace over the $SU(5)$ {\bf 24} matrices, one finds that the theory
describes two real 6-component scalar fields with vacuum field strength
$\sqrt{15} v_1$ and $\sqrt{15} v_2$. If we choose $v_1=v_2$ for example,
then the dimensionless parameter $\epsilon = 8 \pi^2 G\phi_0^2$
governing the magnitude of cosmological perturbations is related to the
unification scale $M_X$ by
\eqn\ecsc{\eqalign{
{M_X^2 \over M^2_{\rm pl} }  &= \epsilon { 5 \alpha_{\rm GUT}\over 24 \pi}\cr
}}
If we set $\epsilon = 10^{-4}$, which is the value
of the field strength required for the $SU(2)$ theory to fit the COBE
microwave anisotropy \PST, then we deduce that $M_X \approx 4 \times
10^{16}$ GeV, consistent with the values obtained by requiring
unification and a sufficiently long proton lifetime in Section 6.
Conversely, from the requirement that $M_X = 10^{15-16}$ GeV, one sees that
$\epsilon $ is constrained to be within an order of magnitude of
the value required by COBE. As far as we are aware,
there is no other theoretical framework which comes this
close to a prediction of the level of primordial density perturbations
based on particle physics considerations alone.

\centerline{\bf Conclusions}

In this paper we have examined in some detail the possibility
of a fundamental symmetry relating the three families of elementary
particles. We would be the first to acknowledge that the scheme
we have explored is  simplistic, and makes no attempt to explain
why such a symmetry should exist. The fact that the symmetry
is in the form of a direct product $G_{\rm fam} \times G_{\rm GUT}$ is
 ugly - the gauge and global symmetries are not unified. We
 have done nothing to ameliorate the heirarchy problem, and have made
 no attempt to include supersymmetry or indeed gravity -
  we might well be criticised for ignoring possible violations
of global symmetries by  quantum gravity (see e.g. \ref\hol{
 R. Holman, S. Hsu, E. Kolb, R. Watkins and
L. Widrow, Phys. Rev. Lett., {\bf 69} (1992) 1489.;
 M. Kamionkowski
and J. March-Russell, Phys. Rev. Lett., {\bf 69} (1992) 1485.}).

Nevertheless we have made a case for a simple family symmetry group
- $SU(3)$ - and argued that it has to be a global symmetry. We have shown
in detail that a renormalisable symmetry breaking Higgs potential
can produce vevs with sufficient parameters to match the measured
fermion masses and quark mixing matrix. We have begun to explore
the low energy phenomenology of the theory - the rich low energy
Higgs sector - and shown how flavour changing neutral currents
might be avoided while remaining in the weak coupling regime.
We have pointed out how the simplest $SU(5)$
theory simplifies when it is embedded in $SO(10)$, and
shown how unification of the coupling constants is rather economically
achieved.
Finally we showed how the global symmetry breaking at the GUT scale
leads to the production of cosmic texture with the correct
symmetry breaking scale to produce structure in the universe.
While the ideas in this paper might be criticised as being
naive, they cannot be criticised for not being testable!

There are several further developments of this work which we
believe could be fruitful.

a) The detailed phenomenology of the eighteen (!) low energy
 electroweak
doublets should be examined. In particular it would be interesting to
know in general which of the
 charged or neutral Higgs bosons would be the easiest
to detect. The consequences of $CP$ violation in the Higgs sector
should also be investigated - the neutron and electron dipole moments,
and the baryon asymmetry produced at the electroweak phase transition.
Of course the main problem is that the parameter space is huge,
but it would be interesting to know whether there are any generic
predictions.

b) Full cosmological simulations of the $SU(3)$ nonlinear sigma model
described in Section 8 may be performed to compute structure formation
and cosmic microwave anisotropies in the theory.

c) The $SO(10)$ theory should be constructed in detail. Predictions
of relations between neutrino masses and mixing angles may be
possible.

d) The theory given here produces stable magnetic monopoles,
which, in the absence of inflation, are cosmologically disastrous.
Nevertheless it may be possible that for some range of parameters
and temperatures the
finite temperature potential has a minimum without a $U(1)$ unbroken
gauge group factor. As Kibble and Weinberg have recently
argued \ref\KW{T.W.B. Kibble and E. Weinberg, Phys. Rev. {\bf D 43}
(1991) 3188.}, in this case
magnetic monopoles would either never be formed at all, or
might be connected by strings and disappear by the Langacker-Pi mechanism.

e) One might search for a fundamental origin of the $SU(3)$ symmetry
invoked here. The fermion kinetic term has an accidental $SU(3)$ global
symmetry, and if one adopted the `technicolor' approach to symmetry
breaking i.e. insisted that there should be no fundamental scalar fields,
it is conceivable that one might preserve this as an exact symmetry
of the theory, spontaneously broken by fermion bilinears
as happens in massless $QCD$.
The Higgs fields we need can all be obtained as bilinears in
fermion fields in the {\bf 16} of $SO(10)$, plus the adjoint representation,
which arises most naturally as the extra space components of a
gauge field in Kaluza Klein theoires.

\centerline{Acknowledgements}
Most of this work was done at Imperial College, London. We would like to
thank A. Albrecht, M. Hindmarsh, K. Rajagopal, D. Spergel and F. Wilczek
for discussions.
The work of MJ and NT was partially  supported by the SERC (UK),
NSF contract PHY90-21984,  the Alfred P. Sloan Foundation, and
the David and Lucile Packard Foundation.

\centerline{\bf Appendix }

    Consider the potential of the ({\bf 24},{\bf 3})s. We want to show that we
can
get the minimum in $\eq$. As in our analysis of the potential
above we will not attempt a general solution but simply
choose certain terms and values of the couplings.

Consider the terms
\eqn\er{\eqalign{-{\mu^2 \over 2}tr(\Sigma_a \Sigma_a^\dagger)
+ {\alpha_1 \over 4}tr(\Sigma_a\Sigma_a^\dagger\Sigma_b^\dagger\Sigma_b)
+ {\alpha_2 \over 4}tr(\Sigma_a \Sigma_a^\dagger)tr(\Sigma_b\Sigma_b^\dagger)
+ {\alpha'_2 \over 4}tr(\Sigma_a\Sigma_b)tr(\Sigma_a^\dagger\Sigma_b^\dagger)}}
where the traces are now over the $SU(5)$ indices. We expand explicitly
in the six hermitian {\bf 24}s in each complex
vector ({\bf 24},{\bf 3}). For example the first term gives
\eqn\err{\eqalign{-{\mu^2 \over 2}\Sigma tr(A_a^2+B_a^2) }}
where we have substituted $\Sigma_a=A_a+iB_a$ and $A_a^{\dagger}=A_a$,
$B_a^{\dagger}=B_a$. Doing the same for every term we get a sum of terms
which consists of two parts:

(i) groups of terms which are the same as the potential of a single
real {\bf 24} i.e.
\eqn\ers{\eqalign{-{\mu^2 \over 2}trA^2 + {\alpha_1 \over 4}trA^4
                 + {\alpha_2 \over 4}(trA^2)^2  }}
which has a stationary point at $A=0$ and at
$A=diag(1,1,1,-{3 \over 2},-{3 \over2})$ if
\eqn\es{\eqalign{v^2 = {4\mu^2 \over {7\alpha_1 + 30\alpha_2}}}}

(ii) remaining terms which are stationary at $\eq$ simply because they
involve more than one of the six real matrices in the decomposition, and are
at least quadratic in any matrix contributing to a given term.

	Thus $\er$ is stationary at $\eq$ if
\eqn\ess{\eqalign{v^2 = {4\mu^2 \over {7\alpha_1 + 30(\alpha_2+\alpha'_2)}}}}
When we expand in perturbations about $\eq$ we find that
the only massless modes are the Goldstone modes generated by action of
the symmetry group. All the other perturbations
can be shown to have positive masses provided $0<{13 \over 60}\alpha_1
<|{\alpha'_2}| <\alpha_2$ (and $\alpha'_2<0)$.

When we now consider this potential for each of the two ({\bf 24},{\bf 3})s
and couple them together with
the positive definite terms
\eqn\et{\eqalign{
tr(\Sigma^1_a\Sigma^{2 \dagger}_a)tr(\Sigma^2_b\Sigma^{1 \dagger}_b
)\qquad
  & (\perp\quad {\rm in}\quad SU(3)) \cr
tr(\Sigma^1_a\Sigma^{1 \dagger}_a)tr(\Sigma^2_b \Sigma^{2 \dagger}_b)
  - tr(\Sigma^1_a \Sigma^{2\dagger}_b)tr(\Sigma^2_b\Sigma^{1 \dagger}_b)
                       \qquad &(\parallel\quad {\rm in}\quad SU(5))\cr }}
we get the required minimum. One can check easily that these coupling
terms $\et$ are sufficient to completely break the symmetry as desired.

\listrefs
\bye